\documentclass[journal]{IEEEtran}

\usepackage{abstract}
\usepackage{amsmath,amssymb,amsfonts}
\usepackage{authblk}
\usepackage[labelfont=bf]{caption}
\usepackage{enumitem}
\usepackage{fancyhdr}
\usepackage{graphicx}
\usepackage{layouts}
\usepackage{mathrsfs}
\usepackage[numbers,sort&compress]{natbib}
\usepackage{siunitx}
\usepackage{xcolor}
\usepackage[commandnameprefix=ifneeded,final]{changes}
\makeatletter
\@namedef{Changes@AuthorColor}{magenta}
\colorlet{Changes@Color}{magenta}
\makeatother

\usepackage{hyperref}
\hypersetup{
    colorlinks,
    citecolor=blue,
    filecolor=blue,
    linkcolor=blue,
    urlcolor=blue
}

\newcommand{\tss}[1]{\textsuperscript{#1}}

\title{\LARGE Bayesian and Monte Carlo approaches to estimating uncertainty for the measurement of the bound-state $\beta^-$ decay of \tss{205}Tl\tss{81+}}

\author[1,2,3,a]{G.~Leckenby\thanks{\tss{a} email: leckenby@lp2ib.in2p3.fr}}
\author[4,b]{M.~Trassinelli\thanks{\tss{b} email: martino.trassinelli@cnrs.fr}}
\author[5,6,7]{R.~J.~Chen}
\author[5,8,9]{R.~S.~Sidhu}
\author[5]{J.~Glorius}
\author[5]{M.~S.~Sanjari}
\author[5,10,11,c]{Yu.~A.~Litvinov\thanks{\tss{c} email: y.litvinov@gsi.de}}
\author[5]{M.~Bai}
\author[5,{\textdagger}]{F.~Bosch\thanks{\tss{\textdagger} Deceased}}
\author[5,12]{C.~Brandau}
\author[5,13]{T.~Dickel}
\author[1,14]{I.~Dillmann}
\author[5,15]{D.~Dmytriiev}
\author[16]{T.~Faestermann}
\author[5,17]{O.~Forstner}
\author[5]{B.~Franczak}
\author[5,13,{\textdagger}]{H. Geissel}
\author[16]{R.~Gernh\"auser}
\author[5,7]{B.~S.~Gao}
\author[1]{C.~J.~Griffin}
\author[5]{A.~Gumberidze}
\author[5]{E.~Haettner}
\author[5]{R.~He\ss}
\author[5,12]{P.-M.~Hillenbrand}
\author[16,{\textdagger}]{P.~Kienle}
\author[19]{W.~Korten}
\author[5]{Ch.~Kozhuharov}
\author[5]{N.~Kuzminchuk}
\author[5]{S.~Litvinov}
\author[5,10]{E.~B.~Menz}
\author[5]{T.~Morgenroth}
\author[5]{C.~Nociforo}
\author[5,{\textdagger}]{F.~Nolden}
\author[5]{N.~Petridis}
\author[5]{U.~Popp}
\author[5]{S.~Purushothaman}
\author[20,21]{R.~Reifarth}
\author[5,11,12]{C.~Scheidenberger}
\author[5]{U.~Spillmann}
\author[5]{M.~Steck}
\author[5,17,18]{Th.~St\"ohlker}
\author[22]{Y.~K.~Tanaka}
\author[5]{S.~Trotsenko}
\author[5,16]{L.~Varga}
\author[7]{M.~Wang}
\author[5]{H.~Weick}
\author[8]{P.~J.~Woods}
\author[23]{T.~Yamaguchi}
\author[7]{Y.~H.~Zhang}
\author[5]{J.~W.~Zhao}

\affil[1]{Department of Physical Sciences, TRIUMF, Vancouver, BC V6T 2A3, Canada}
\affil[2]{Department of Physics and Astronomy, University of British Columbia, Vancouver, BC V6T 1Z1, Canada}
\affil[3]{LP2i Bordeaux, CNRS, Universit\'e de Bordeaux, F-33170 Gradignan, France}
\affil[4]{Institut des NanoSciences de Paris, CNRS, Sorbonne Universit\'e, F-75005 Paris, France}
\affil[5]{GSI Helmholtzzentrum f{\"u}r Schwerionenforschung GmbH, Planckstra{\ss}e 1, D-64291 Darmstadt, Germany}
\affil[6]{Max-Planck-Institut f{\"u}r Kernphysik, Saupfercheckweg 1, D-69117 Heidelberg, Germany}
\affil[7]{CAS Key Laboratory of High Precision Nuclear Spectroscopy, 730000 Lanzhou, China} 
\affil[8]{School of Physics and Astronomy, University of Edinburgh, EH9 3FD, UK}
\affil[9]{School of Mathematics and Physics, University of Surrey, Guildford, GU2 7XH, UK}
\affil[10]{Institut f\"ur Kernphysik, Universit\"at zu K\"oln,  Z{\"u}lpicher Str. 77, D-50937 K\"oln, Germany}
\affil[11]{Helmholtz Forschungsakademie Hessen f{\"u}r FAIR (HFHF), GSI Helmholtzzentrum f{\"u}r Schwerionenforschung GmbH, 64291 Darmstadt, Germany}
\affil[12]{I. Physikalisches Institut, Justus-Liebig Universit{\"a}t Gie{\ss}en, D-35392 Gie{\ss}en, Germany}
\affil[13]{II. Physikalisches Institut, Justus-Liebig Universit{\"a}t Gie{\ss}en, D-35392 Gie{\ss}en, Germany}
\affil[14]{Department of Physics and Astronomy, University of Victoria, Victoria, BC V8P 5C2, Canada}
\affil[15]{Deutsches Elektronen Synchrotron DESY, D-22603 Hamburg, Germany}
\affil[16]{Physik Department E12, Technische Universität M\"{u}nchen, D-85748 Garching, Germany}
\affil[17]{Friedrich-Schiller-Universit\"at Jena, D-07743 Jena, Germany}
\affil[18]{Helmholtz-Institut Jena, Fraunhoferstra{\ss}e, D-07743 Jena, Germany}
\affil[19]{IRFU, CEA, Universit\'{e} Paris-Saclay, F-91191 Gif-sur-Yvette, France}
\affil[20]{J.W. Goethe Universit{\"a}t, D-60438 Frankfurt, Germany}
\affil[21]{Los Alamos National Laboratory, Los Alamos, NM-87545, USA}
\affil[22]{High Energy Nuclear Physics Laboratory, RIKEN, Wak{\=o} 351-0198, Japan}
\affil[23]{Saitama University, Saitama 338-8570, Japan}

\begin{document}
\twocolumn[\begin{@twocolumnfalse} 
\maketitle
	\begin{abstract}
		The measurement of the bound-state $\beta$ decay of \tss{205}Tl at the Experimental Storage Ring (ESR) at GSI, Darmstadt, has recently been reported with substantial impact on the use of \tss{205}Pb as an early Solar System chronometer and the low-energy measurement of the solar neutrino spectrum via the LOREX project. Due to the technical challenges in producing a high-purity \tss{205}Tl\tss{81+} secondary beam, a robust statistical method needed to be developed to estimate the variation in the contaminant \tss{205}Pb\tss{81+} produced \added{in the fragmentation reaction, which was subsequently transmitted and stored in the ESR}. Here we show that Bayesian and Monte Carlo methods produced comparable estimates for the contaminant variation, each with unique advantages and challenges given the complex statistical problems for this experiment. We recommend the adoption of such methods in future experiments that exhibit unknown statistical fluctuations.

        \vspace{10pt}
	\end{abstract}
\end{@twocolumnfalse}]
\saythanks


\section{Introduction}\label{sec:intro}
Bound-state $\beta^-$ decay ($\beta_b$ decay) has proven to be a rare but consequential decay mode in astrophysical events because it has the potential to vastly alter the decay properties of certain nuclei~\cite{daudel_sur_1947,bahcall_theory_1961}. In this decay mode, the $\beta$ electron is created in a bound-state of the daughter nucleus, which can significantly enhance the $Q$ value of the decay for very high charge states. Takahashi~\emph{et al.}~\cite{takahashi_bound-state_1987} were the first to catalogue key isotopes where bound-state $\beta$ decay had a large impact on the decay rate, and this was soon backed up by the first measurements at the Experimental Storage Ring (ESR) at GSI Helmholtzzentrum f\"ur Schwerionenforschung in Darmstadt, Germany~\cite{jung_first_1992,bosch_observation_1996,ohtsubo_simultaneous_2005,Kurcewicz-2010}. The technical challenges involved in storing millions of fully-stripped ions for several hours mean that, presently, the ESR is the only facility capable of directly measuring bound-state $\beta$ decay\,\cite{Litvinov-2011, Bosch-2013, Steck-2020}.

Recently, we reported on the first measurement of the $\beta_b$ decay of \tss{205}Tl\tss{81+}~\cite{leckenby_high-temperature_2024,sidhu_bound-state_2024}. This measurement was crucial in establishing the weak decay rates of \tss{205}Tl and \tss{205}Pb in the core of Asymptotic Giant Branch (AGB) stars, the nucleosynthetic site for the cosmochronometer \tss{205}Pb that is valuable in early Solar System studies. Additionally, the $\beta_b$-decay rate of \tss{205}Tl\tss{81+} was required for the Lorandite Experiment (LOREX), which aims to use the Tl-bearing mineral lorandite to achieve the lowest-threshold measurement of the solar neutrino spectrum, sensitive to $E_{\nu_e}\geq53~\unit{keV}$\,\cite{Pavicevic-2018}. 

An accurate estimation of uncertainties on experimental measurements are paramount in nuclear astrophysics, especially when trying to determine whether discrepancies originate from nuclear data or astrophysical modelling. For the $\beta_b$ decay of \tss{205}Tl\tss{81+}, our experimental half-life is 4.7 times longer than the values presently used in the astrophysical community~\cite{takahashi_beta-decay_1987}, although agrees with modern shell-model calculations~\cite{ogawa_shell-model_1988,warburton_first-forbidden_1991,xiao_calculations_2024}, so an accurate uncertainty was crucial in order to assess the impact of the experiment. The analysis we presented in Refs.~\cite{leckenby_high-temperature_2024,sidhu_bound-state_2024}, with further details in the theses~\cite{leckenby_exotic_2025,sidhu_measurement_2021}, was quite complex involving four corrections and a fit with experimentally measured parameters. This analysis demanded careful handling of the correlations between data points and a fit to a model with uncertain parameters, which required methods more sophisticated than the traditional $\chi^2$~minimisation. Furthermore, we identified in Refs.~\cite{leckenby_high-temperature_2024,sidhu_bound-state_2024} that the measured uncertainties on the corrections, which we refer to as the ``raw uncertainties,'' do not sufficiently describe the scatter of the data. To address both these issues, we developed a Monte Carlo error propagation that included a method for self-consistently estimating the ``missing uncertainty'' and including it in the determination of the model parameters, namely the decay-rate of \tss{205}Tl\tss{81+}.

This Monte Carlo method extends a Frequentist framework to deal with experimentally determined parameters and missing uncertainty, however, Bayesian methods also naturally handle these issues in a self-consistent way. In this work we compare the two approaches for the decay-rate estimation.
Additionally, we explore how the Bayesian approach can efficiently handle outlier data points, which we had to exclude manually in our original analysis. 
To further investigate the origin of the missing uncertainty, we also present a new, dedicated analysis of the statistical uncertainties. Due to the presence of beam losses, this analysis is not trivial and it has been further developed here to take into account the Poisson nature of the recorded signals.
In Sec.~\ref{sec:experiment}, we briefly introduce the experimental method. \added{Since the experimental method and analysis corrections have been described elsewhere, we refer the interested reader to the provided references for further details.} The Monte Carlo uncertainty estimation method is then presented in Sec.~\ref{sec:monte-carlo}, with Poisson counting statistics described in Sec.~\ref{sec:uncertainty}. The alternative Bayesian analysis in then presented Sec.~\ref{sec:bayesian}, with a discussion on the comparison of the methods in Sec.~\ref{sec:discussion} and a conclusion in Sec.~\ref{sec:conclusion}.

\section{Experimental Method}\label{sec:experiment}
In our experimental analysis, the sources of uncertainty for each datum depend on the method for producing and detecting the ions. To measure the $\beta_b$ decay of \tss{205}Tl, ions need to be stored in the storage ring in the fully-stripped charge state as only $\beta_b$~decay to the K shell of the daughter nucleus is energetically allowed. To do so, a secondary \tss{205}Tl\tss{81+} beam was created via projectile fragmentation of a primary \tss{206}Pb\tss{67+} beam using the entire accelerator chain at GSI. The main contaminant of concern was \tss{205}Pb\tss{81+}, as it has only a 31.1(5)~keV mass difference compared to \tss{205}Tl\tss{81+} and is also the $\beta_b$-decay daughter product, which directly confounds the decay signal. To minimise this contamination, the Fragment Separator (FRS) was used, where an aluminium energy degrader produced a spatial separation at the exit of the FRS via the $B\rho$--$\Delta E$--$B\rho$ method~\cite{geissel_gsi_1992}. With this setting, \tss{205}Pb\tss{81+} contamination was reduced to $\sim0.1\%$.

Roughly $10^6$ \tss{205}Tl\tss{81+} ions were accumulated in the ESR from over 100 injections from the FRS. These ions were then stored in the ring for various periods up to 10 hours to accumulate \tss{205}Pb\tss{81+} decay products. Ions in the ring where monitored with a non-destructive, resonant Schottky detector~\cite{Nolden-2011,sanjari_resonant_2013}. Because of the very small mass difference, \tss{205}Tl\tss{81+} and \tss{205}Pb\tss{81+} ions revolve on nearly identical orbits in the storage ring, so the \tss{205}Pb\tss{81+} decay daughters were mixed within the \tss{205}Tl\tss{81+} parent beam. To count the \tss{205}Pb\tss{81+} decay daughters, the stored \tss{205}Tl/Pb\tss{81+} beam was interacted with an argon gas-jet target with a density of $\sim10^{12}~\unit{atoms/cm^2}$ that stripped off the bound electron revealing the \tss{205}Pb daughters in the 82+ charge state that could then be counted with the Schottky detectors. The $N_{\mathrm{Pb}^{81+}}/N_{\mathrm{Tl}^{81+}}$ ratio at the end of the storage period could subsequently be determined from the measured $N_{\mathrm{Pb}^{82+}}/N_{\mathrm{Tl}^{81+}}$ ratio after gas stripping\,\cite{Chen-EPJA}. 

\added{The $N_{\mathrm{Pb}^{81+}}/N_{\mathrm{Tl}^{81+}}$ ratio at the end of the storage period is given by
\begin{equation}\label{eqn:ratio_calc}
    \frac{N_\text{Pb}(t_s)}{N_\text{Tl}(t_s)}=\frac{SA_\text{Pb}(t_s)}{SA_\text{Tl}(t_s)} \, \frac{1}{SC(t_s)} \, \frac{1}{RC} \, \frac{\epsilon_\text{Tl}(t_s)}{\epsilon_\text{Pb}(t_s)} \, \frac{\sigma_\text{str}+\sigma_\text{rec}}{\sigma_\text{str}}.
\end{equation}}
\replaced{This equation features four corrections that were required to successfully extract the ratio:}{The method for extracting the $N_{\mathrm{Pb}^{82+}}/N_{\mathrm{Tl}^{81+}}$ ratio at the end of the storage period required four corrections:}
\begin{enumerate}\itemsep0em
    \item a saturation correction \added{$SC(t_s)$} to account for a mismatched amplifier in the Schottky DAQ this experiment;
    \item a resonance correction \added{$RC$} to account for the resonance response of the Schottky cavity at different revolution frequencies;
    \item the interaction efficiency \added{$\epsilon(t_s)$} to determine the extent to which both species interacted with the gas target; and
    \item a charge-changing cross section ratio \added{$(\sigma_\text{str}+\sigma_\text{rec})/\sigma_\text{str}$} that accounts for \tss{205}Pb\tss{81+} ions lost to electron recombination rather than stripping.
\end{enumerate}
This experimental protocol is covered in Refs.~\cite{leckenby_high-temperature_2024,sidhu_bound-state_2024} with extensive details on the analysis corrections given in the theses~\cite{leckenby_exotic_2025,sidhu_measurement_2021}, \added{including validation and quantification of each correction to the total uncertainty. The intermediate and result data for this experiment are publicly available at Ref.~\cite{leckenby_measurement_2024} and the analysis script is also available at Ref.~\cite{leckenby_measurement_2024-1}.}

It has been established in previous $\beta_b$-decay experiments~\cite{jung_first_1992}, and derived explicitly in Ref.~\cite{sidhu_measurement_2021}, that as the storage time increases, the daughter/parent ratio will increase pseudo-linearly according to:
\begin{align}
    \begin{split} 
        \frac{N_\mathrm{Pb}(t_s)}{N_\mathrm{Tl}(t_s)}=\frac{\lambda_{\beta_b}}{\gamma}t_s\big[1&+\tfrac{1}{2}(\lambda^\mathrm{loss}_\mathrm{Tl}-\lambda^\mathrm{loss}_\mathrm{Pb})t_s\big]\\
        &+\frac{N_\mathrm{Pb}(0)}{N_\mathrm{Tl}(0)}\exp\big[(\lambda^\mathrm{loss}_\mathrm{Tl}-\lambda^\mathrm{loss}_\mathrm{Pb})t_s\big],
    \end{split}
    \label{eqn:bsbd-growth}
\end{align}  
where $N_X$ is the number of \tss{205}Pb or \tss{205}Tl ions, $t_s$ is the storage time, $\lambda_{\beta_b}$ is the $\beta_b$-decay rate of \tss{205}Tl\tss{81+}, and $\gamma=1.429(1)$ is the Lorentz factor for conversion into the laboratory frame. The initial \tss{205}Pb\tss{81+} contamination, written here as $N_\mathrm{Pb}(0)$, must be scaled by the storage loss rates $\lambda^\mathrm{loss}_X$, which are slightly different for Pb and Tl. Thus, Eq.~\eqref{eqn:bsbd-growth} is the physical model that describes data.

It is important to note that the uncertainties in the Schottky intensities and the interaction efficiencies were determined for each storage period and are thus statistically independent, whilst the uncertainties in the saturation correction, resonance correction, the charge-changing cross section, and the storage losses were determined globally for the entire data set and are thus entirely correlated between data points. As a result, it was not straightforward to incorporate those uncertainties into the fit of Eq.~\eqref{eqn:bsbd-growth}. Furthermore, the fit itself contains parameters with experimental uncertainties that need to be included. This cannot be handled by a simple $\chi^2$ minimisation, as the $\chi^2$ assumption assumes that the data points are statistically independent. To handle these correlations between the data and the fit, and to take into account the uncertainties of the parameters of Eq.~\eqref{eqn:bsbd-growth}, the Monte Carlo error propagation method was chosen.

\section{Monte Carlo approach}\label{sec:monte-carlo}
Monte Carlo (MC) error propagation is a well-established method that is used when the uncertainty distribution or the physical model is too complicated to compute \added{the uncertainties analytically~\cite{jcgm_evaluation_2008,possolo_invited_2017,cox_use_2006}}.\deleted{ or with specific constraints.} It simulates ``$m$ runs'' of the experiment, where the underlying uncertainty distributions are randomly sampled and then that sampled data set is used \replaced{to fit}{ in} the physical model. In our case, the underlying uncertainty distributions are the experimental uncertainties coming from the \added{thermal and electronic noise in the} Schottky detectors and \added{the evaluated uncertainties of the experimental} corrections\added{, which are mostly Gaussian (details in Ref.~\cite{leckenby_exotic_2025})}. \added{Because some of the experimental corrections are applied globally to all 16 data points (e.g. resonance correction, charge-changing cross section, etc.), their uncertainties are correlated and this correlation is included in the MC simulation because these corrections are sampled only once and applied to all data points for a given run.} \replaced{Additionally, we sampled the uncertainties of the experimentally determined parameters of the physical model.}{ as well as the uncertain model parameters.} The physical model is Eq.~\eqref{eqn:bsbd-growth} and to fit a sampled data set with this equation we just used least squares to give a best fit value for the decay rate $\lambda_{\beta_b}$. The distribution in the final model parameters is then representative of the underlying uncertainty distributions and the correlations between uncertainties, and even the ``nuisance'' model parameters are naturally handled by the MC simulation. 

\begin{figure}[h!]
    \centering
    \includegraphics[width=0.85\linewidth]{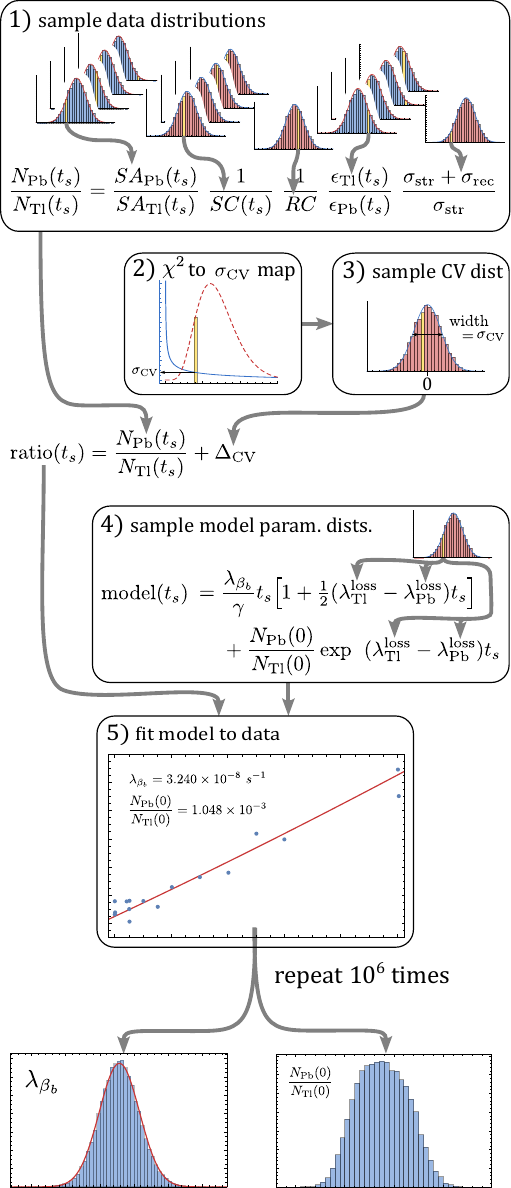}
    \caption{\added{A flow chart outlining the MC method following five steps described in the text. The yellow bin schematically suggests that this value was the sampled value for that MC run. The distributions in blue are uncorrelated between the 16 measurements, whilst distributions in red are correlated. Multiple distributions stacked indicates that this individual values were used for this variable for each of the 16 measurements, whereas a single distribution indicates a single value for all measurements.}}
    \label{fig:MC_flow}
\end{figure}

\added{Fig.~\ref{fig:MC_flow} presents a flow chart that summarises our implementation of the MC method, which can be broken down into five steps:}
\begin{enumerate}\itemsep0em
    \item Randomly sample each measured quantity (Schottky intensities, corrections, etc.) and calculate the corrected $N_\mathrm{Pb}/N_\mathrm{Tl}$ ratio \added{with Eq.~\eqref{eqn:ratio_calc}} to produce 16 data points.
    \item Randomly sample the $\chi^2(\nu=14)$ distribution and determined a value for $\sigma_\mathrm{CV}$ for that run (see discussion below for motivation).
    \item Add a randomly sampled value to each data point from a Gaussian distribution with mean zero and standard deviation $\sigma_\mathrm{CV}$ to simulate the contamination variation.
    \item \added{Randomly sample the uncertain model parameters ($\lambda^\text{loss}$ rates) to construct a model from Eq.~\eqref{eqn:bsbd-growth} for that MC run.}
    \item Use least squares to fit \added{the sampled model}, \deleted{using randomly sampled values for the experimentally measured model parameters (e.g. $\lambda^\mathrm{loss}$),} which gives a best fit value for $\lambda_{\beta_b}$ and $N_\mathrm{Pb}(0)/N_\mathrm{Tl}(0)$. 
\end{enumerate}
\added{This method is then repeated $10^6$ times to produce a distribution of final model parameters. Our code for the MC method is also available publicly in Ref.~\cite{leckenby_measurement_2024-1}.}

The strengths of the MC method is that it is straightforward to apply to an arbitrarily complicated analysis whilst simultaneously handling the correlations between the inputs, even if the input uncertainties have highly unusual distributions, which was the primary motivation for our analysis. The weakness is that it is a numerical method so the precision of the model parameter distributions is determined by the number of MC runs $m$, which can be quite constraining if the analysis is computationally expensive. Typically $m>10^4$ is considered sufficient, and for our relatively simple model fit, we were able to achieve a MC accuracy of 0.02\% with $m=10^6$ MC runs.

\begin{figure}[t]
    \centering
    \includegraphics[width=\linewidth]{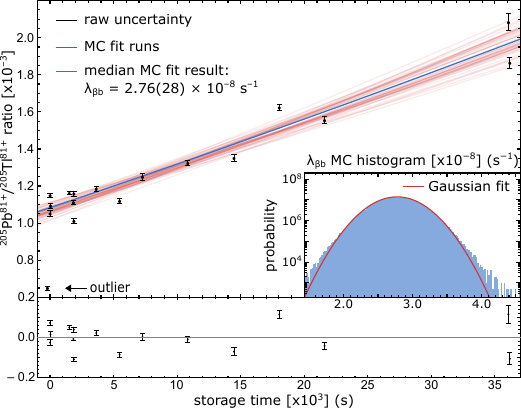}
    \caption{The \tss{205}Pb/\tss{205}Tl ratios for all storage periods, plotted with the ``raw uncertainties'' from the corrections, alongside a sample 100 MC fits (red) and the median MC fit result (blue) \added{that represents the median value of the MC histogram of the final model parameters}. The residuals (lower panel) highlight that some statistical variation is missing to explain the scatter of the data. The inset shows the MC histogram for $\lambda_{\beta_b}$ with the Gaussian fit used to extract the model parameter distribution.}
    \label{fig:MC_results}
\end{figure}

Figure~\ref{fig:MC_results} shows the experimental data, \replaced{with each data point representing the ratio derived at the end of a storage run}{ with the data points} with the \added{error bars representing the} ``raw statistical error'' derived from the median and $\pm1\sigma$ intervals of the MC simulation for that \replaced{ratio}{point}. Also plotted are the best fit results from 100 sample MC runs with the inset showing the \replaced{histogram of the MC best fit results of the $\beta_b$ decay rate.}{ MC histogram for the decay rate.}

Looking to the residuals in Fig.~\ref{fig:MC_results}, it is clear that the ``raw uncertainty'' from the corrections cannot explain the scatter of the data, indicating not all of the experimental uncertainty was accounted for. In particular, the $\chi^2$ of the data in Fig.~\ref{fig:MC_results} is 303, whereas the 95\% confidence interval for a $\chi^2$ distribution with 14 degrees of freedom is $[6.6,\,23.7]$. The first remedy for missing uncertainty should always be the careful examination of the data and the identification and quantification of the unaccounted for uncertainty. For this purpose, we surveyed a list of possible sources  of systematic effects, and the details are provided in \S3.3.1 of Ref.~\cite{leckenby_exotic_2025}. Notably, Poisson ``counting statistics'' cannot account for all the missing uncertainty, providing variation at roughly the 3\% level (see Sec~\ref{sec:uncertainty} for a deeper discussion). By exclusion of all other possibilities, we concluded that this variation most likely arose from fluctuations in the field strengths of the FRS magnets between storage periods. Given the FRS selected a $>3\sigma$ tail from the \tss{205}Pb fragmentation distribution, it is likely that such fluctuations could produce the missing uncertainty of $\sim6\%$ in the \tss{205}Pb contamination level that is observed. 

Since there was no method to measure this contamination variation from the FRS as it was perfectly confounded with our signal, the missing uncertainty must be quantified by analysing the scatter of the data. Whilst bootstrapping by random re-sampling of the data is a self-consistent method with minimal assumptions that was used in earlier parts of this analysis, our simulations have shown that it is only accurate if more than 50 data points are available. With only 16 data points, we were forced to turn to other methods. 

We first considered the well-known Birge ratio method~\cite{Birge1932}. Discussed at length in our recent publication Ref.~\cite{Trassinelli2025}, the Birge ratio is used to globally increase all error bars by a factor $\Tilde{\sigma}_i=R_B\,\sigma_i$ with
\begin{equation}
    R_B=\sqrt{\frac{1}{n-k}\sum_i\left(\frac{y_i-f[x_i|\vec{\alpha}]}{\sigma_i}\right)^2}=\sqrt{\frac{\chi^2}{n-k}},
\end{equation}
where $n$ is the number of data points $(x_i,y_i)$ being evaluated and $k$ the number of parameters $\vec{\alpha}$ of the model. This has the effect of globally inflating the error bars until the $\chi^2=n-k$. However, inflating the error bars maintains the relative size of the error bars, which in our case underweights the long storage time measurements that contain the most signal from $\beta_b$ decay. Specifically, the uncertainty coming from the corrections is much smaller for the short storage times, whilst the contamination variation from the FRS should affect all data points equally. For this reason, we chose to add additional uncertainty $\sigma_\mathrm{CV}$ in quadrature rather than inflate the raw uncertainties, yielding a $\chi^2$ of:
\begin{equation}
    \chi^2=\sum_i\frac{(y_i-f[x_i|\vec{\alpha}])^2}{\sigma_{i,\mathrm{stat}}^2+(\exp[(\lambda^\mathrm{loss}_\mathrm{Tl}-\lambda^\mathrm{loss}_\mathrm{Pb})t_s]\times\sigma_\mathrm{CV})^2}.
\end{equation}
Note that the contamination variation $\sigma_\mathrm{CV}$ must be multiplied by the storage loss factor $\exp[(\lambda^\mathrm{loss}_\mathrm{Tl}-\lambda^\mathrm{loss}_\mathrm{Pb})t_s]$ to account for the evolution of the contamination with storage time.

\begin{figure}[t]
    \centering
    \includegraphics[width=\linewidth]{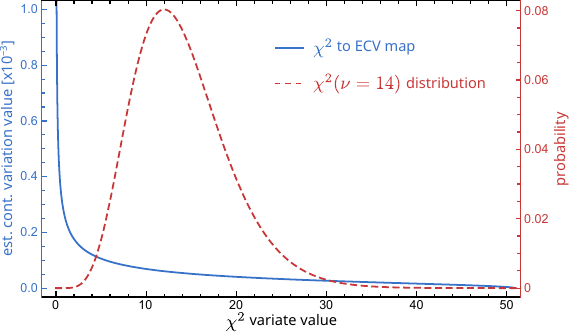}
    \caption{The mapping (in blue) between the $\chi^2$ variate and the estimated contamination variation $\sigma_\mathrm{CV}$ required to achieve that value for our data and model. The $\chi^2(\nu=14)$ probability distribution is overlayed (in red) to guide the eye in regard to how often a given value was sampled.}
    \label{fig:chisqr_mapping}
\end{figure}

The Birge ratio also breaks down for small numbers of data points because the $\chi^2$ distribution is no longer constraining. Consider our case with $n-k=14$ where the 95\% confidence interval of the associated $\chi^2$ distribution is $[6.6,\,23.7]$. Using the Birge ratio to achieve a final $\chi^2$ value of 14 is not well motivated over other values in this range, which has lead to criticism of the method over the years. However, this problem can be naturally solved by the Monte Carlo propagation method because for each MC run a different target value for the $\chi^2$ can be chosen. Specifically, we constructed a mapping between the $\sigma_\mathrm{CV}$ value and the minimum $\chi^2$ value for that $\sigma_\mathrm{CV}$ value. Using the mapping, we could then randomly sample a $\chi^2$ variate for that MC run and determine how much contamination variation to add to the data for that MC simulation run. This allows us to naturally incorporate the uncertainty with which the data can constrain the missing statistical variation that is identified by the $\chi^2$ goodness of fit test, solving a key weakness of the Birge ratio using the unique framework of the Monte Carlo method. The mapping between the value of $\chi^2$ and $\sigma_\mathrm{CV}$ is plotted in Fig.~\ref{fig:chisqr_mapping} overlayed with the $\chi^2(\nu=14)$ probability distribution that was sampled. 

The histogram of best fit values for $\lambda_{\beta_b}$---shown in the inset of Fig.~\ref{fig:MC_results}---gives the final uncertainty distribution for the parameter: $\lambda_{\beta_b}=2.76(28)\times10^{-8}~\unit{s^{-1}}$, as reported in refs.~\cite{leckenby_high-temperature_2024,sidhu_bound-state_2024}. \added{This distribution includes the raw statistical error as well as the correlated systematic errors and the estimated contamination variation.}

Another well known weakness of the $\chi^2$ minimisation method is the sensitivity to possible outliers \cite{vonderLinden,Rousseeuw,Trassinelli2025}.
A single outlier can dramatically increase the calculated $\chi^2$, thus increasing the estimated missing uncertainty as well as biasing the result. We saw this effect in our data analysis as a result of the outlier tagged in the bottom-left of Fig.~\ref{fig:MC_results}. When included in the analysis, it increased the slope of the fit and doubled the uncertainty giving a final distribution of $\lambda_{\beta_b}=3.13(47)\times10^{-8}~\unit{s^{-1}}$. This outlier was $6.7\sigma$ away from the best fit prediction, even with the missing uncertainty included, so we discarded this data point for the above MC analysis. This case highlights a further weakness of using the $\chi^2$ to estimate missing uncertainty because it must be paired with judicious data selection in order to provide reasonable results, which can be slow, painstaking work when done carefully.

\section{Impact of counting statistics} \label{sec:uncertainty}
It was determined fairly quickly that the uncertainty arising from counting a finite number of decays, otherwise known as ``counting statistics'', could not explain the missing variation in our data (producing $\sim3\%$ variation whereas 6\% was missing).\deleted{ and so we expected its impact to be negligible.} Since the variation from the counting statistics is automatically included in the estimation of the contamination variation in the MC method, \added{we did not differentiate between these two sources in the analysis reported in Refs.~\cite{leckenby_high-temperature_2024,sidhu_bound-state_2024,leckenby_exotic_2025,sidhu_measurement_2021}}. This sufficed for an accurate determination of $\lambda_{\beta_b}$, however, as we shall see, an in depth treatment of counting statistics is necessary to compare the MC method with a Bayesian analysis.

In a typical decay counting experiment, the impact of counting statistics is accounted for via the Poisson uncertainties on the counting bins or ideally via the Maximum Likelihood Estimator where each data point is considered individually. The $\beta_b$-decay measurements are atypical because the ions cannot be identified at the time of decay since they are mixed with the parents ions due to the small $Q$ value of the decay, and thus ions must be counted at the end of the storage period. The number of ions that decay during a given storage period follow a Poisson distribution, but furthermore, the ions are affected by beam losses and the initial ion population in this experiment arose from projectile fragmentation, both of which are also Poisson processes. Fortunately, the sum of two independent Poisson variables $X\sim P(\lambda)$ and $Y\sim P(\eta)$ is itself a Poisson variable with $X+Y\sim P(\lambda+\eta)$, so the Poisson variation from all of these processes is just determined by the number of ions at any given time. It is worth noting that whilst the parent and daughter populations are 100\% correlated for radioactive decay, the loss constants $\lambda^\mathrm{loss}$ are three orders-of-magnitude larger than $\lambda_{\beta_b}$, so the variance of the $N_\mathrm{Tl}$ and $N_\mathrm{Pb}$ populations are not correlated in this experiment.

\begin{figure}[t]
    \centering
    \includegraphics[width=\linewidth]{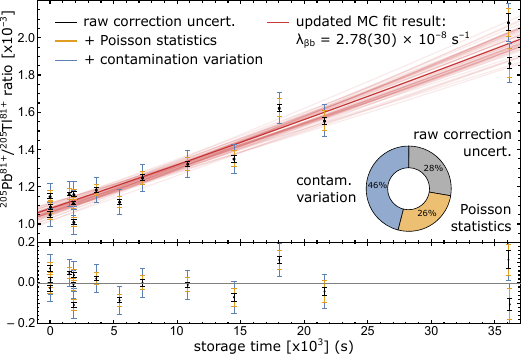}
    \caption{The updated MC method with error bars indicating the variance included for Poisson statistics and updated contamination variation. \added{The example 100 MC fits are again shown in red with the median fit result used to calculate the residuals.} The pie chart shows the contribution of each source to the total uncertainty, highlighting that the dominant uncertainty for the experiment remains the contamination variation from the FRS.}
    \label{fig:poisson}
\end{figure}

For the first two $\beta_b$-decay experiments~\cite{jung_first_1992,bosch_observation_1996}, the fully-stripped $\beta_b$-decay daughters were counted with particle detectors so the number of ions was known immediately. For \tss{205}Tl\tss{81+}, we chose to measure the ratio of Schottky noise power densities---proportional to the ion number according to the Schottky theorem---to avoid introducing systematic errors when calibrating the Schottky detectors. Whilst the Schottky noise power density has its own uncertainty, it does not account for variance introduced by the Poisson statistics. The Poisson uncertainty in the ratio $R=N_\mathrm{Pb}/N_\mathrm{Tl}$ is just $\sigma_R=\sqrt{R/N_\mathrm{Tl}}$ (see App.~\ref{app:poisson_errs} for derivation). Using this expression, we can add Poisson uncertainty to each data point using the MC method in the same way we added the contamination variation. The results are shown in Fig.~\ref{fig:poisson}. As can be seen from the residuals, Poisson statistics alone are not enough to account for the missing uncertainty, although they do explain a third of the \added{unaccounted-for} variation.\deleted{, which is not negligible.} However, as expected, the final result from an updated MC analysis is $\lambda_{\beta_b}=2.78(30)\times10^{-8}~\unit{s^{-1}}$, almost identical to the published result, indicating the Poisson variation had been included in the contamination variation. Despite this, \replaced{explicitly modeling}{including} Poisson statistics is essential when comparing the MC method with the Bayesian analysis because we are primarily interested in the two method's behaviour when estimating the missing uncertainty.

\section{Fully Bayesian approach}\label{sec:bayesian}

The fully Bayesian approach, like the previously described MC method, considered the correct evaluation of the likelihood taking into account the uncertainty of the parameters of the implemented model.
For this purpose, each experimental datum was evaluated as a Gaussian probability distribution centred at the mean value
with a standard deviation equal to the associated uncertainty.
However, differently from the method described in the previous section, it does not simply minimise the likelihood but efficiently explores the allowed parameter space taking into account possible multimodalities.
Unlike the Monte Carlo method, the dependence between data is ignored.
The posterior probabilities are obtained by evaluating the likelihood function over the parameter space using the nested sampling method \cite{Skilling2004,Ashton2022} and, more specifically, the \textsc{nested\_fit} code \cite{Trassinelli2017b,Trassinelli2019,Maillard2023}.
The nested sampling is a recursive search algorithm using an ensemble of sampling points that evolves during the algorithm execution.
It is a method normally used for Bayesian model selection, where, in addition to finding the maximum of the likelihood function, a probability for each model is evaluated from the integral over the parameter space of the model (via the calculation of the Bayesian evidence).
\added{Here, however, the \textsc{nested\_fit} code is not used for the Bayesian model selection but only for the powerful exploration features of the nested sampling and the capability of \textsc{nested\_fit} to treat multimodal solutions and to use non-standard likelihoods.}

\begin{figure}
    \centering
    \includegraphics[width=0.8\linewidth]{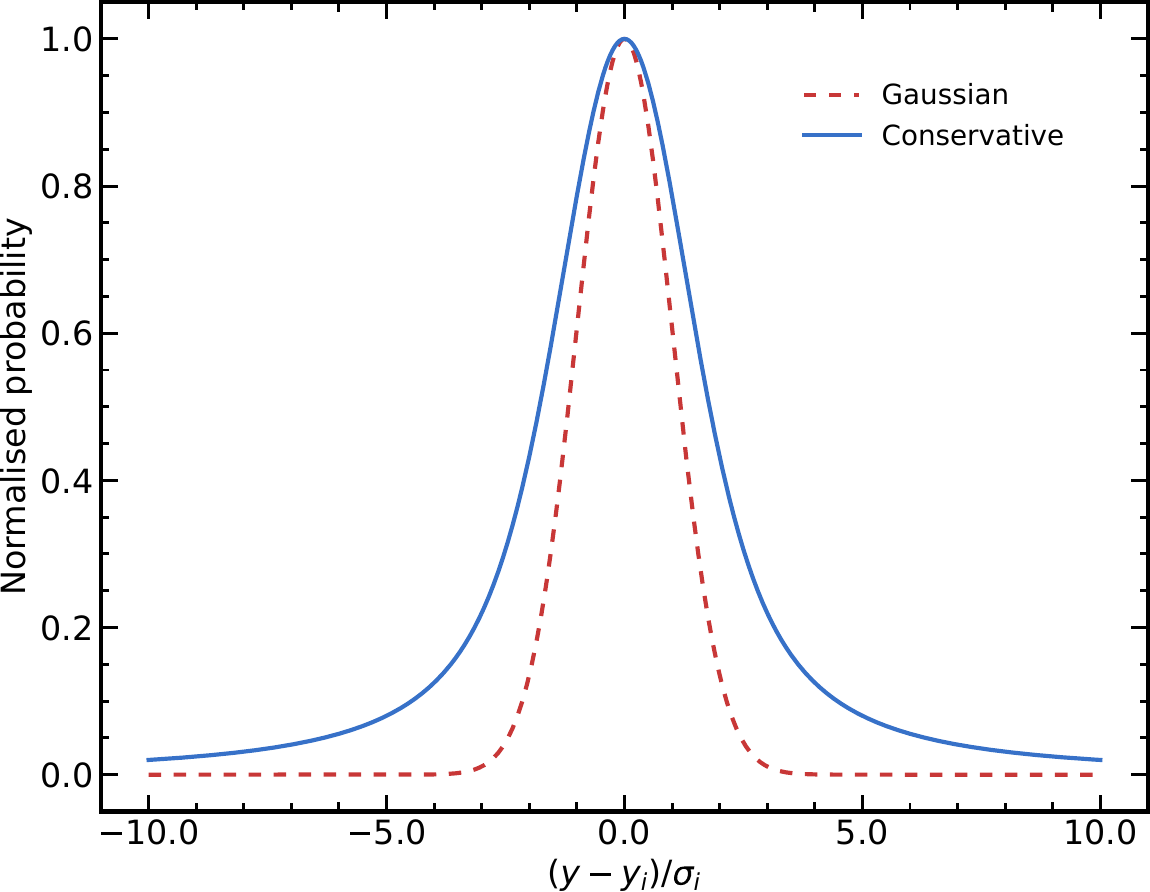}
    \caption{Standard Gaussian distribution compared to the distribution derived by the Bayesian conservative method for each single data point $y_i$ with standard deviation $\sigma_i$.}
    \label{fig:profiles}
\end{figure}

The standard procedure builds the likelihood function by considering a normal (Gaussian) distribution of the experimental data with respect to the model value, with mean equal to the experimental value $y_i$ and the standard deviation equal to the value of the error bar $\sigma_i$.
The maximisation of the likelihood thus corresponds to the minimisation of the associated $\chi^2$ value.
Alternatively, we consider also the approach proposed by Sivia \cite{Sivia}, referred here as the \emph{conservative approach}, for a robust analysis.
For each data point, the error bar value is instead considered as a lower bound on the real unknown uncertainty $\sigma'_i$ that includes unevaluated systematic contributions.
After marginalisation over the possible values of $\sigma'_i$ (considering a modified version of non-informative Jeffreys prior), \added{for each datum} the corresponding distribution is no longer a normal distribution, but now features lateral tails decreasing as $1/(y - y_i)^2$ (see Fig.~\ref{fig:profiles}), 
\added{where $y=y_i-f[x_i|\vec{\alpha}]$ is the theoretical expected value for the modelling function $f[x_i|\vec{\alpha}]$.
More precisely, the final likelihood function $L$ is now given by
\begin{equation}
    L = \prod_i  \frac {\sigma_i} {\sqrt{2 \pi}} \left[ \frac {1 - e^{\frac{(y_i-f[x_i|\vec{\alpha}])^2}{2 \sigma^2_i}}}{(y_i-f[x_i|\vec{\alpha}])^2} \right].
\end{equation}
}
Because of the slower drop-off at extreme values of $y$, such a distribution is naturally more tolerant to outliers and, more generally, to inconsistency between error bars and data dispersion.

\begin{figure}
    \centering
    \includegraphics[width=\linewidth]{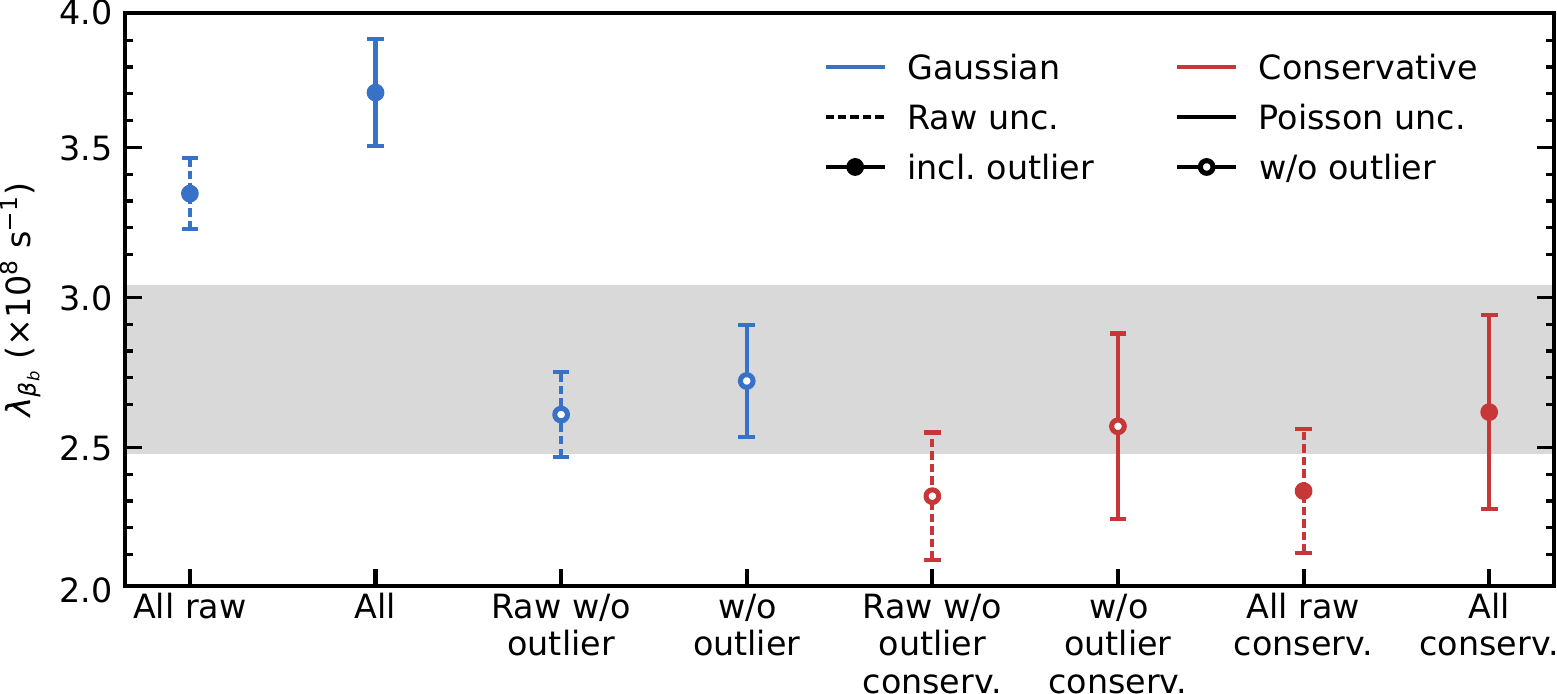}
    \caption{Comparison between the different statistical analysis. The grey band correspond to the reported value in \cite{leckenby_high-temperature_2024}. The other approaches, indexed in the legend with three possible variations, are described in the text.}
    \label{fig:comparison}
\end{figure}

By using the \textsc{nested\_fit} code, different estimates for the $\beta_b$ decay constant are extracted by analysing the same data as in Ref.~\cite{leckenby_high-temperature_2024} (``raw uncertainties'') or with the Poisson statistical additional contribution discussed in Sec.~\ref{sec:uncertainty}.
We consider in addition the entire available data set (``all'') with and without the outlier in the bottom-left of Fig.~\ref{fig:MC_results} that was excluded from the MC method. 
Finally, we considered both standard Gaussian and conservative distributions.
The different results are presented in Fig.~\ref{fig:comparison}.
\added{For all different cases, like the Monte Carlo method, the uncertainties of the parameters $\lambda^\mathrm{loss}_\mathrm{Tl}$ and $\lambda^\mathrm{loss}_\mathrm{Pb}$ have been taken into account by considering Gaussian prior probabilities with the parameter's uncertainty as standard deviation (and its nominal value as mean).
Due to the very small relative uncertainty ($10^{-3}$), no priors were considered for $\gamma$, which is considered as exact.
The slice sampling \cite{Neal2003} is used for the search of the sampling points in \textsc{nested\_fit} because to its superior performance with respect to other methods (like random walk, see \cite{Maillard2023}).
Typical computation times were in the order of a few seconds on a standard laptop PC.
}

As expected, the presence of outlier significantly introduces an important bias in the determination of the decay constant when the standard likelihood is implemented, even considering the uncertainty evaluations from the Poisson statistics of the ions.
Once the outlier is excluded, the obtained decay value is consistent with the MC method reported in Ref.~\cite{leckenby_high-temperature_2024}, where the uncertainties were globally increased to justify the residual scattering. 

When the conservative method is adopted, no difference can be noted if the outlier is considered or not, demonstrating the robustness of the method.
However, the corresponding uncertainties are larger than the other cases because of the conservative method, which assumes that there are missing systematic uncertainty contributions for each data point.
It is remarkable that for the conservative method, even if the error bars of the raw data are underestimated, the final evaluation of the decay constant is not far from the one obtained by the elimination of the outlier and the artificial enhancing of the error bars using the Monte Carlo method discussed in Sec.~\ref{sec:monte-carlo} (the grey band in Fig.~\ref{fig:comparison}).
The use of a more tolerant probability distribution (Fig.~\ref{fig:profiles}) leads, in fact, to a final uncertainty that depends on the data dispersion.
This is not completely the case for standard $\chi^2$ minimisation methods.
See Ref.~\cite{Trassinelli2025} for a more extended discussion on this aspect, but applied for weighted averages.
Between the different approaches, the preferable option is the one with the least working hypotheses considering the best of our possible knowledge at present.
This corresponds to considering the entire set with the Poisson statistics uncertainty resulting in the parameter estimate of $\lambda_{\beta_b}=2.62(32)\times10^{-8}~\unit{s^{-1}}$.

\section{Discussion}\label{sec:discussion}







As one can see in the previous sections, the different robust statistical approaches comfortingly give very similar results, even when the statistical uncertainty is underestimated.

The Monte Carlo method is a robust technique for handling underestimated uncertainties in small data sets whilst respecting the uncertainty in the ability of the data to constrain the missing uncertainty, namely the fluctuation in the $\chi^2$ distribution. Simultaneously, it takes into account correlation in the data and uncertainties in the nuisance parameters of the model making it quite a flexible tool. An outstanding weakness is that outlier detection must be done manually to ensure an accurate estimate of the missing uncertainty.

The Bayesian method is simpler in the sense that it requires less hypotheses about the data. Taking advantage of the Bayesian formulation, uncertain parameters can be represented with experimentally determined prior distributions, and then that uncertainty is naturally included when they are integrated out in the final parameter estimation. The implemented missing uncertainty estimation is also optimised to handle outliers, making it a very flexible tool as well. Correlation between the data points, however, cannot be considered in the current likelihood construction. This appears to not be limiting in this analysis as the uncertainty is dominated by the uncorrelated contamination variation and Poisson counting statistics.

The two methods produce remarkably consistent results, as shown in Fig.~\ref{fig:comparison}, especially when considering they handle the missing uncertainty in a fundamentally different way. The MC method globally adds the same additional uncertainty to all data points so the whole data set is scaled together. This is an appropriate choice if the unknown variation has affected all data points equally and has the advantage that no data point is treated preferentially, albeit reducing the constraining power of the more accurate points. The Bayesian method, on the other hand, treats the missing uncertainty on each data point individually. This optimises the Bayesian method to handle outliers, but does allow for situations where the minimisation will choose between two, otherwise equal data points to get a better fit.

\section{Conclusions}\label{sec:conclusion}
The estimation of unknown statistical error is a common problem in experimental physics, and whilst every effort should be made to understand the uncertainties of the experiment, sometimes sources of uncertainty need to be estimated from the data. The method for estimating this missing uncertainty should be as accurate and unbiased as possible. One of the most common methods for estimating missing uncertainty has been the Birge ratio. However, we have explained how it is biased towards the most probable $\chi^2$ of the data and does not take into account the full range of possibilities for model parameters that are compatible with the data. We have presented two methods that provide an alternative to the Birge ratio. The Monte Carlo method takes advantage of the fact that it uses multiple trials, so the $\chi^2$ value that is replicated can be varied with each trial. The Bayesian method takes advantage of the fact that Bayesian analysis can self-consistently allow the data to constrain the missing uncertainty. As a result, both of these methods improve on the traditional Birge ratio and more reliably represent the uncertainty of measured parameters.

The application of multiple analysis methods also provides a very useful assessment on the variation in our final value and uncertainty. We have shown here that across a broad range of analysis methods, our final result for the $\beta_b$ decay rate is reliable with a very reproducible uncertainty. A correct treatment of the Poisson statistics improves the coherence of the results \added{by decomposing the missing uncertainty into its Poisson and contamination variation components explicitly.} \deleted{this reinforces the importance of a good estimation of uncertainties.} \added{It is also reassuring to note that this dedicated analysis resulted in only a 0.7\% or $0.07\sigma$ increase on the final value reported in Refs.~\cite{leckenby_high-temperature_2024,sidhu_bound-state_2024}.} Once counting statistics are included, which is essential for a fair comparison, the MC and Bayesian methods agree on the reported value within $0.5\sigma$ \added{and provide almost identical estimates for the size of the parameter uncertainty. This highlights the remarkable robustness of our reported half-life value considering how different the methods are.} 

The presence of the outlier (bottom-left in Fig.~\ref{fig:MC_results}) also tested the method's ability to handle outliers. The Bayesian method clearly is better suited to handle outliers because the missing uncertainty is evaluated individually for each data point, \added{allowing individual outliers to be inflated without penalising the entire data set.} The fact that the method can handle both the outlier and the missing uncertainty out-of-the-box and get a very similar result to the MC method, which required months of work to perfect, demonstrates its versatility and reliability in analysing challenging data. As a result, we recommend the use of the open-source \textsc{nested\_fit} code \cite{Trassinelli2017b,Trassinelli2019,Maillard2023} for future experimental analysis. For those who want an alternative to Bayesian methods, which can be sometimes computational costly, or where strong correlations in the data exist, we also recommend the Monte Carlo method for easy error propagation and a natural mechanism for estimating missing uncertainty.

\section*{Acknowledgements}
The results presented here are based on the experiment G-17-E121, which was performed at the FRS-ESR facilities at the GSI Helmholtzzentrum f{\"u}r Schwerionenforschung, Darmstadt (Germany) in the frame of FAIR Phase-0.
This project has received funding from the European Research Council (ERC) under the European Union's Horizon 2020 research and innovation programme (Grant Agreement No. 682841 ``ASTRUm'' and No. 654002 ``ENSAR2'').
The research of G.~Leckenby, I.~Dillmann, and C.~Griffin was funded by the Canadian Natural Sciences and Engineering Research Council (NSERC) via the grant SAPIN-2019-00030. 
J.~Glorius, M.~S.~Sanjari, Yu.~A.~Litvinov and C.~Brandau acknowledge support by the State of Hesse within the Research Cluster ELEMENTS (Project ID 500/10.006). E.~Menz and Yu.~A.~Litvinov acknowledge support by the project ``NRW-FAIR", a part of the programme ``Netzwerke 2021", an initiative of the Ministry of Culture and Science of the State of North Rhine-Westphalia.
R.~Gernh\"auser acknowledges support by the Excellence Cluster ORIGINS from the German Research Foundation DFG (Excellence Strategy EXC-2094–390783311).

\vfill\pagebreak
\bibliographystyle{myunsrtnat}
\bibliography{CPC_Errors_paper}

\begin{thebibliography}{39}
\providecommand{\natexlab}[1]{#1}
\providecommand{\url}[1]{\texttt{#1}}
\expandafter\ifx\csname urlstyle\endcsname\relax
  \providecommand{\doi}[1]{doi: #1}\else
  \providecommand{\doi}{doi: \begingroup \urlstyle{rm}\Url}\fi

\bibitem[Daudel et~al.(1947)Daudel, Jean, and Lecoin]{daudel_sur_1947}
Daudel, R., Jean, M. and Lecoin, M.
\newblock \emph{J. Phys. Rad.}, \href{http://dx.doi.org/10.1051/jphysrad:0194700808023800}{\textbf{8}:\penalty0 238--243}, (1947).

\bibitem[Bahcall(1961)]{bahcall_theory_1961}
Bahcall, J.N.
\newblock \emph{Phys. Rev.}, \href{http://dx.doi.org/10.1103/PhysRev.124.495}{\textbf{124}:\penalty0 495--499}, (1961).

\bibitem[Takahashi et~al.(1987)Takahashi, Boyd, Mathews, and Yokoi]{takahashi_bound-state_1987}
Takahashi, K. \emph{et~al}.
\newblock \emph{Phys. Rev. C}, \href{http://dx.doi.org/10.1103/PhysRevC.36.1522}{\textbf{36}:\penalty0 1522--1528}, (1987).

\bibitem[Jung et~al.(1992)Jung, Bosch, Beckert, Eickhoff, Folger, Franzke, Gruber, Kienle, Klepper, Koenig, Kozhuharov, Mann, Moshammer, Nolden, Schaaf, Soff, Sp{\"a}dtke, Steck, St{\"o}hlker, and S{\"u}mmerer]{jung_first_1992}
Jung, M. \emph{et~al}.
\newblock \emph{Phys. Rev. Lett.}, \href{http://dx.doi.org/10.1103/PhysRevLett.69.2164}{\textbf{69}:\penalty0 2164--2167}, (1992).

\bibitem[Bosch et~al.(1996)Bosch, Faestermann, Friese, Heine, Kienle, Wefers, Zeitelhack, Beckert, Franzke, Klepper, Kozhuharov, Menzel, Moshammer, Nolden, Reich, Schlitt, Steck, Stolker, and Winkler]{bosch_observation_1996}
Bosch, F. \emph{et~al}.
\newblock \emph{Phys. Rev. Lett.}, \href{http://dx.doi.org/10.1103/PhysRevLett.77.5190}{\textbf{77}:\penalty0 5190--5193}, (1996).

\bibitem[Ohtsubo et~al.(2005)Ohtsubo, Bosch, Geissel, Maier, Scheidenberger, Attallah, Beckert, Beller, Boutin, Faestermann, Franczak, Franzke, Hausmann, Hellstr{\"o}m, Kaza, Kienle, Klepper, Kluge, Kozhuharov, Litvinov, Matos, M{\"u}nzenberg, Nolden, Novikov, Portillo, Radon, Stadlmann, Steck, St{\"o}hlker, S{\"u}mmerer, Takahashi, Weick, Winkler, and Yamaguchi]{ohtsubo_simultaneous_2005}
Ohtsubo, T. \emph{et~al}.
\newblock \emph{Phys. Rev. Lett.}, \href{http://dx.doi.org/10.1103/PhysRevLett.95.052501}{\textbf{95}:\penalty0 052501}, (2005).

\bibitem[Kurcewicz et~al.(2008)Kurcewicz, Bosch, Geissel, Litvinov, Winckler, Beckert, Beller, Boutin, Brandau, Chen, Dimopoulou, Essel, Fabian, Faestermann, Fragner, Franzke, Haettner, Hausmann, Hess, Kienle, Kn{\"o}bel, Kozhuharov, Litvinov, Maier, Mazzocco, Montes, Musumarra, Nociforo, Nolden, Patyk, Plass, Prochazka, Reda, Reuschl, Scheidenberger, Steck, St\"ohlker, Sun, Takahashi, Torilov, Trassinelli, Weick, and Winkler]{Kurcewicz-2010}
Kurcewicz, J. \emph{et~al}.
\newblock \emph{Acta Phys. Polon. B}, \href{https://www.actaphys.uj.edu.pl/R/39/2/501/pdf}{\textbf{41}:\penalty0 525--536}, (2008).

\bibitem[Litvinov and Bosch(2011)]{Litvinov-2011}
Litvinov, Y.A. and Bosch, F.
\newblock \emph{Rep. Prog. Phys.}, \href{http://dx.doi.org/10.1088/0034-4885/74/1/016301}{\textbf{74}:\penalty0 016301}, (2011).

\bibitem[Bosch et~al.(2013)Bosch, Litvinov, and St{\"o}hlker]{Bosch-2013}
Bosch, F., Litvinov, Y.A. and St{\"o}hlker, T.
\newblock \emph{Prog. Part. Nucl. Phys.}, \href{http://dx.doi.org/10.1016/j.ppnp.2013.07.002}{\textbf{73}:\penalty0 84 -- 140}, (2013).

\bibitem[Steck and Litvinov(2020)]{Steck-2020}
Steck, M. and Litvinov, Y.A.
\newblock \emph{Progress in Particle and Nuclear Physics}, \href{http://dx.doi.org/https://doi.org/10.1016/j.ppnp.2020.103811}{\textbf{115}:\penalty0 103811}, (2020).

\bibitem[Leckenby et~al.(2024{\natexlab{a}})Leckenby, Sidhu, Chen, Mancino, Sz{\'a}nyi, Bai, Battino, Blaum, Brandau, Cristallo, Dickel, Dillmann, Dmytriiev, Faestermann, Forstner, Franczak, Geissel, Gernh{\"a}user, Glorius, Griffin, Gumberidze, Haettner, Hillenbrand, Karakas, Kaur, Korten, Kozhuharov, Kuzminchuk, Langanke, Litvinov, Litvinov, Lugaro, Mart{\'i}nez-Pinedo, Menz, Meyer, Morgenroth, Neff, Nociforo, Petridis, Pignatari, Popp, Purushothaman, Reifarth, Sanjari, Scheidenberger, Spillmann, Steck, St{\"o}hlker, Tanaka, Trassinelli, Trotsenko, Varga, Vescovi, Wang, Weick, Yag{\"u}e~Lop{\'e}z, Yamaguchi, Zhang, and Zhao]{leckenby_high-temperature_2024}
Leckenby, G. \emph{et~al}.
\newblock \emph{Nature}, \href{http://dx.doi.org/10.1038/s41586-024-08130-4}{\textbf{634}:\penalty0 321--326}, (2024){\natexlab{a}}.

\bibitem[Sidhu et~al.(2024)Sidhu, Leckenby, Chen, Mancino, Neff, Litvinov, Mart{\'i}nez-Pinedo, Amthauer, Bai, Blaum, Boev, Bosch, Brandau, Cvetkovi{\'c}, Dickel, Dillmann, Dmytriiev, Faestermann, Forstner, Franczak, Geissel, Gernh{\"a}user, Glorius, Griffin, Gumberidze, Haettner, Hillenbrand, Kienle, Korten, Kozhuharov, Kuzminchuk, Langanke, Litvinov, Menz, Morgenroth, Nociforo, Nolden, Pavi{\'c}evi{\'c}, Petridis, Popp, Purushothaman, Reifarth, Sanjari, Scheidenberger, Spillmann, Steck, St{\"o}hlker, Tanaka, Trassinelli, Trotsenko, Varga, Wang, Weick, Woods, Yamaguchi, Zhang, Zhao, Zuber, and Collaborations]{sidhu_bound-state_2024}
Sidhu, R.S. \emph{et~al}.
\newblock \emph{Phys. Rev. Lett.}, \href{http://dx.doi.org/10.1103/PhysRevLett.133.232701}{\textbf{133}:\penalty0 232701}, (2024).

\bibitem[Pavi{\'c}evi{\'c} et~al.(2018)Pavi{\'c}evi{\'c}, Amthauer, Cvetkovi{\'c}, Boev, Pejovi{\'c}, Henning, Bosch, Litvinov, and Wagner]{Pavicevic-2018}
Pavi{\'c}evi{\'c}, M.K. \emph{et~al}.
\newblock \emph{Nucl. Instr. Meth. A}, \href{http://dx.doi.org/10.1016/j.nima.2018.03.039}{\textbf{895}:\penalty0 62 -- 73}, (2018).

\bibitem[Takahashi and Yokoi(1987)]{takahashi_beta-decay_1987}
Takahashi, K. and Yokoi, K.
\newblock \emph{At. Data Nucl. Data Tables}, \href{http://dx.doi.org/10.1016/0092-640X(87)90010-6}{\textbf{36}:\penalty0 375--409}, (1987).

\bibitem[Ogawa and Arita(1988)]{ogawa_shell-model_1988}
Ogawa, K. and Arita, K.
\newblock \emph{Nucl. Instrum. Methods Phys. Res., Sect. A}, \href{http://dx.doi.org/10.1016/0168-9002(88)90169-6}{\textbf{271}:\penalty0 280--285}, (1988).

\bibitem[Warburton(1991)]{warburton_first-forbidden_1991}
Warburton, E.K.
\newblock \emph{Phys. Rev. C}, \href{http://dx.doi.org/10.1103/PhysRevC.44.233}{\textbf{44}:\penalty0 233--260}, (1991).

\bibitem[Xiao and Wang(2024)]{xiao_calculations_2024}
Xiao, Y. and Wang, L.J.
\newblock \emph{Phys. Rev. C}, \href{http://dx.doi.org/10.1103/PhysRevC.110.054308}{\textbf{110}:\penalty0 054308}, (2024).

\bibitem[Leckenby(2025)]{leckenby_exotic_2025}
Leckenby, G.
\newblock \emph{Exotic decay measurements at the {Experimental} {Storage} {Ring} for neutron capture processes}.
\newblock {PhD},  (\href{https://dx.doi.org/10.14288/1.0447637}{University of British Columbia}, 2025).

\bibitem[Sidhu(2021)]{sidhu_measurement_2021}
Sidhu, R.S.
\newblock \emph{Measurement of the bound-state beta decay of bare {205Tl81}+ ions at the {ESR}}.
\newblock {PhD},  (\href{https://doi.org/10.11588/heidok.00030275}{Heidelberg University}, 2021).

\bibitem[Geissel et~al.(1992)Geissel, Armbruster, Behr, Brünle, Burkard, Chen, Folger, Franczak, Keller, Klepper, Langenbeck, Nickel, Pfeng, Pfützner, Roeckl, Rykaczewski, Schall, Schardt, Scheidenberger, Schmidt, Schröter, Schwab, Sümmerer, Weber, Münzenberg, Brohm, Clerc, Fauerbach, Gaimard, Grewe, Hanelt, Knödler, Steiner, Voss, Weckenmann, Ziegler, Magel, Wollnik, Dufour, Fujita, Vieira, and Sherrill]{geissel_gsi_1992}
Geissel, H. \emph{et~al}.
\newblock \emph{Nuclear Instruments and Methods in Physics Research Section B: Beam Interactions with Materials and Atoms}, \href{http://dx.doi.org/10.1016/0168-583X(92)95944-M}{\textbf{70}:\penalty0 286--297}, (1992).

\bibitem[Nolden et~al.(2011)Nolden, H{\"u}lsmann, Litvinov, Moritz, Peschke, Petri, Sanjari, Steck, Weick, Wu, Zang, Zhang, and Zhao]{Nolden-2011}
Nolden, F. \emph{et~al}.
\newblock \emph{Nucl. Instr. Meth. A}, \href{http://dx.doi.org/10.1016/j.nima.2011.06.058}{\textbf{659}:\penalty0 69 -- 77}, (2011).

\bibitem[Sanjari et~al.(2013)Sanjari, H{\"u}lsmann, Nolden, Schempp, Wu, Atanasov, Bosch, Kozhuharov, Litvinov, Moritz, Peschke, Petri, Shubina, Steck, Weick, Winckler, Zang, and Zhao]{sanjari_resonant_2013}
Sanjari, M.S. \emph{et~al}.
\newblock \emph{Phys. Scripta}, \href{http://dx.doi.org/10.1088/0031-8949/2013/T156/014088}{\textbf{T156}:\penalty0 014088}, (2013).

\bibitem[Chen et~al.(2025)]{Chen-EPJA}
Chen, R.J. \emph{et~al}.
\newblock \emph{Eur. Phys. J. A}, \textbf{in press}, (2025).

\bibitem[Leckenby et~al.(2024{\natexlab{b}})Leckenby, Sidhu, Chen, Bai, Blaum, Brandau, Dickel, Dillmann, Dmytriiev, Faestermann, Forstner, Franczak, Giessel, Gernhäuser, Glorius, Griffin, Gumberidze, Haettner, Hillenbrand, Korten, Kozhuharov, Kuzminchuk, Langanke, Litvinov, Litvinov, Menz, Morgenroth, Nociforo, Petridis, Popp, Purushothaman, Reifarth, Sanjari, Scheidenberger, Spillmann, Steck, Stöhlker, Tanaka, Trotsenko, Varga, Wang, Weick, Yamaguchi, Zhang, and Zhao]{leckenby_measurement_2024}
Leckenby, G. \emph{et~al}.
\newblock Measurement of the bound-state beta decay of {205Tl}(81+): intermediate and result data.
\newblock  (\href{https://zenodo.org/records/11556665}{Zenodo}, 2024).

\bibitem[Leckenby et~al.(2024{\natexlab{c}})Leckenby, Sidhu, Chen, Szányi, Dillmann, Gernhäuser, Glorius, Griffin, Litvinov, Lugaro, Sanjari, and Yagüe~Lopéz]{leckenby_measurement_2024-1}
Leckenby, G. \emph{et~al}.
\newblock Measurement of the bound-state beta decay of {205Tl}(81+): analysis scripts and figures.
\newblock  (\href{https://zenodo.org/records/11560338}{Zenodo}, 2024).

\bibitem[{BIPM, IEC, IFCC, ILAC, ISO, IUPAC, IUPAP and OIML}(2008)]{jcgm_evaluation_2008}
{BIPM, IEC, IFCC, ILAC, ISO, IUPAC, IUPAP and OIML}.
\newblock Evaluation of {Measurement} {Data} --- {Supplement} 1 to the {``\emph{{Guide} to the expression of uncertainty in measurement}''} --- {Propagation} of distributions using a {Monte} {Carlo} method.
\newblock \href{https://www.bipm.org/documents/20126/2071204/JCGM_101_2008_E.pdf}{Technical Report JCGM 101:2008}, International Bureau of Weights and Measures (BIPM), (2008).

\bibitem[Possolo and Iyer(2017)]{possolo_invited_2017}
Possolo, A. and Iyer, H.K.
\newblock \emph{Review of Scientific Instruments}, \href{http://dx.doi.org/10.1063/1.4974274}{\textbf{88}:\penalty0 011301}, (2017).

\bibitem[Cox and Siebert(2006)]{cox_use_2006}
Cox, M.G. and Siebert, B.R.L.
\newblock \emph{Metrologia}, \href{http://dx.doi.org/10.1088/0026-1394/43/4/S03}{\textbf{43}:\penalty0 S178}, (2006).

\bibitem[Birge(1932)]{Birge1932}
Birge, R.T.
\newblock \emph{Phys. Rev.}, \href{http://dx.doi.org/10.1103/PhysRev.40.207}{\textbf{40}:\penalty0 207--227}, (1932).

\bibitem[Trassinelli and Maxton(2025)]{Trassinelli2025}
Trassinelli, M. and Maxton, M.
\newblock A minimalistic and general weighted averaging method for inconsistent data.
\newblock  (2025).
\newblock preprint arXiv:2406.08293, submitted to Eur. Phys. J. D.

\bibitem[von~der Linden et~al.(2014)von~der Linden, Dose, and von Toussaint]{vonderLinden}
von~der Linden, W., Dose, V. and von Toussaint, U.
\newblock \emph{Bayesian Probability Theory: Applications in the Physical Sciences}.
\newblock  (Cambridge University Press, 2014).

\bibitem[Rousseeuw and Leroy(2005)]{Rousseeuw}
Rousseeuw, P. and Leroy, A.
\newblock \emph{Robust Regression and Outlier Detection}.
\newblock  (Wiley, 2005).

\bibitem[Skilling(2004)]{Skilling2004}
Skilling, J.
\newblock \emph{AIP Conf. Proc.}, \href{http://dx.doi.org/doi:http://dx.doi.org/10.1063/1.1835238}{\textbf{735}:\penalty0 395--405}, (2004).

\bibitem[Ashton et~al.(2022)Ashton, Bernstein, Buchner, Chen, Csányi, Fowlie, Feroz, Griffiths, Handley, Habeck, Higson, Hobson, Lasenby, Parkinson, Pártay, Pitkin, Schneider, Speagle, South, Veitch, Wacker, Wales, and Yallup]{Ashton2022}
Ashton, G. \emph{et~al}.
\newblock \emph{Nat. Rev. Methods Primers}, \textbf{2}:\penalty0 39, (2022).

\bibitem[Trassinelli(2017)]{Trassinelli2017b}
Trassinelli, M.
\newblock \emph{Nucl. Instrum. Methods B}, \href{http://dx.doi.org/http://dx.doi.org/10.1016/j.nimb.2017.05.030}{\textbf{408}:\penalty0 301--312}, (2017).

\bibitem[Trassinelli(2019)]{Trassinelli2019}
Trassinelli, M.
\newblock \emph{Proceedings}, \href{http://dx.doi.org/http://dx.doi.org/10.3390/proceedings2019033014}{\textbf{33}:\penalty0 14}, (2019).

\bibitem[Maillard et~al.(2023)Maillard, Finocchi, and Trassinelli]{Maillard2023}
Maillard, L., Finocchi, F. and Trassinelli, M.
\newblock \emph{Entropy}, \href{http://dx.doi.org/10.3390/e25020347}{\textbf{25}:\penalty0 347}, (2023).

\bibitem[Sivia and Skilling(2006)]{Sivia}
Sivia, D.S. and Skilling, J.
\newblock \emph{Data analysis: a Bayesian tutorial}.
\newblock  (Oxford University Press, 2006).

\bibitem[Neal(2003)]{Neal2003}
Neal, R.M.
\newblock \emph{Ann. Statist.}, \href{http://dx.doi.org/10.1214/aos/1056562461}{\textbf{31}:\penalty0 705--767}, (2003).

\end{thebibliography}

\appendices

\section{Estimation of Counting Statistics}\label{app:poisson_errs}
To estimate the contribution of counting statistics to our final data points, we want to calculate the variance of our ratio data points arising from the underlying Poisson variables. Noting that the variance of a Poisson distribution is equal to its mean, the definition of the  ratio $R=N_\mathrm{Pb}/N_\mathrm{Tl}$ implies that the variance of $R$ is given by
\begin{align}\label{eqn:rat_variance}
\begin{split}
    \mathrm{Var}(R)&=\,\left(\frac{\partial R}{\partial N_\mathrm{Pb}}\right)^2\mathrm{Var}(N_\mathrm{Pb})+\left(\frac{\partial R}{\partial N_\mathrm{Tl}}\right)^2\mathrm{Var}(N_\mathrm{Pb})\\
    &\qquad+\left(\frac{\partial R}{\partial N_\mathrm{Pb}}\right)\left(\frac{\partial R}{\partial N_\mathrm{Tl}}\right)\mathrm{Cov}(N_\mathrm{Pb},N_\mathrm{Tl})\\
    &=\frac{1}{N_\mathrm{Tl}^2}N_\mathrm{Pb}+\left(-\frac{N_\mathrm{Pb}}{N_\mathrm{Tl}^2}\right)^2N_\mathrm{Tl}\\
    &\qquad-\frac{N_\mathrm{Pb}}{N_\mathrm{Tl}^3}\left(\sqrt{N_\mathrm{Pb}}\right)\left(\sqrt{N_\mathrm{Tl}}\right)\,\mathrm{Corr}(N_\mathrm{Pb},N_\mathrm{Tl})\\
    &=\frac{R}{N_\mathrm{Tl}}+\frac{R^2}{N_\mathrm{Tl}^2}-\frac{R^{3/2}}{N_\mathrm{Tl}}\,\mathrm{Corr}(N_\mathrm{Pb},N_\mathrm{Tl}).
\end{split}
\end{align}
Since $R\ll1$ in our case, we have that $\mathrm{Var}(R)\approx R/N_\mathrm{Tl}$. Note that since we only measured the ratio via Schottky detectors and $N_\mathrm{Tl}$ by the DC current transformer, we have expressed $\mathrm{Var}(R)$ just in terms of these variables.

\end{document}